\documentclass[%
 reprint,
 amsmath,amssymb,
 aps,pre, superscriptaddress
]{revtex4-2}

\usepackage{graphicx}
\usepackage{dcolumn}
\usepackage{bm}
\usepackage{float}
\usepackage{graphicx}

\begin{document}

\preprint{APS/123-QED}

\title{Spin Glass Mapping of the Parallel Minority Game}

\author{Aryan Tyagi}
\affiliation{
School of Computational \& Integrative Sciences,
Jawaharlal Nehru University, New Delhi-110067, India
}%

\author{Soumyaditya Das}%

\affiliation{ 
Department of Physics, SRM University - AP, Andhra Pradesh 522240, India
}%

\author{Soumyajyoti Biswas}%
\email{soumyajyoti.b@srmap.edu.in }
\affiliation{ 
Department of Physics, SRM University - AP, Andhra Pradesh 522240, India
}%

\author{Anirban Chakraborti}
  \email{anirban@jnu.ac.in}
\affiliation{
School of Computational \& Integrative Sciences,
Jawaharlal Nehru University, New Delhi-110067, India
}%

\date{\today}

\begin{abstract}
The parallel minority game (PMG) extends the classical minority game to many choices, with each agent restricted to two predetermined alternatives. In this condition, minimizing the population variance across all choices is a complex combinatorial optimization problem. We show that this minimization is exactly equivalent to finding the ground state of an Ising spin glass in the mean-field limit, i.e., the Sherrington-Kirkpatrick model. By encoding the agent choices as spin variables, the variance becomes a quadratic Hamiltonian with quenched random couplings $J_{ij}$ and random fields $h_i$. This mapping reveals inherent frustration and connects the PMG to the well developed theory of spin glasses, providing a new perspective on the frozen, sub‑optimal configurations observed in stochastic strategies.
\end{abstract}

\maketitle

\section{Introduction}

The minority game (MG) is a paradigmatic model of competitive resource allocation among independent agents, where an odd number of players repeatedly choose between two options, aiming to be in the minority \cite{challet1997,challet1998,challet1999}. Successful strategies can reduce the population fluctuations far below the random choice level, leading to emergent cooperation \cite{challet2001local,sysiaho2004adaptive,chakraborti2011econophysicsII,chakraborti2015}. Stochastic strategies inspired by the Kolkata Paise Restaurant problem have proven particularly efficient, driving the system to a near uniform distribution very quickly \cite{chakrabarti2009,ghosh2010,dhar2011,biswas2012}.

A natural generalization is the \emph{parallel minority game} (PMG) \cite{biswas2021,pmg_vac}. Here, there are $D$ possible locations, but each agent is allowed to switch only between two specific locations, fixed in advance. The choices of different agents overlap, coupling many underlying minority games. As a result, reducing the global population variance $\sigma^2 = (1/D)\sum_l (n_l - N/D)^2$ becomes a much harder optimization problem than in the standard MG. Numerical studies show that even the best stochastic strategies leave a residual variance and eventually freeze into sub‑optimal configurations, reminiscent of spin glasses \cite{tyagi2026,vemula2025}.

In this paper, we demonstrate that the PMG optimization problem maps exactly onto an Ising spin glass \cite{sk_model, sg_pre}. By introducing spin variables $s_i = \pm 1$ to encode each agent's choice, the variance of the occupancies becomes a quadratic energy function with quenched random couplings and fields. The mapping makes explicit the frustration inherent in the PMG, and provides a rigorous foundation for understanding the glassy behavior observed in this system \cite{mezard1987spin} and leaves a scope for exploration of strategy optimization dynamics \cite{Kirkpatrick1983}.

\section{The Parallel Minority Game}

Consider \(N\) agents and \(D\) locations. Each agent \(i\) is permanently assigned two
allowed locations, \(\ell_i^{(1)}\) and \(\ell_i^{(2)}\), chosen uniformly at random from
the \(D\) possibilities. At any moment, agent \(i\) occupies exactly one of these two
sites. The individual objective of every agent is to be in the minority between its two
options: an agent who finds itself in the site with fewer occupants among its two
allowed choices is ``winning''.

Crucially, agents do not coordinate to achieve a uniform global distribution. However,
if all agents manage to place themselves in the minority of their own pair of sites, the
population tends to spread out nearly uniformly across the available locations. A large
number of agents in the minority implies that no site is heavily overcrowded. Therefore,
the global variance
\begin{equation}
E = \sum_{l=1}^{D} \left( n_l - \frac{N}{D} \right)^2 ,
\label{eq:cost}
\end{equation}
where \(n_l\) is the number of agents currently at location \(l\), serves as a natural
cost function that quantifies the deviation from perfect uniformity. For \(N=D\), the
ideal distribution gives \(E=0\). In the following we will reinterpret this cost
function as the energy of a suitably defined spin glass Hamiltonian.

For \(N>D\), some sites inevitably host more than one agent, and even a perfect
local minority strategy cannot eliminate all crowding. Minimising \(E\) then becomes a
non‑trivial combinatorial task. The difficulty stems from two ingredients that are
typical of disordered, frustrated systems:

\begin{itemize}
\item \emph{Quenched disorder.} The allowed pairs \(\{\ell_i^{(1)},\ell_i^{(2)}\}\) are
  fixed throughout the dynamics and are drawn independently for each agent. This
  randomness is the analogue of the random couplings \(J_{ij}\) in spin‑glass models.

\item \emph{Frustration.} A move that improves an agent’s minority status reduces the
  occupancy of one site but necessarily increases the occupancy of the agent’s other
  allowed site. Because many agents share the same sites, a choice that benefits one
  agent may force another into the majority. This competition resembles the frustration
  created by mixed ferromagnetic and antiferromagnetic interactions in Ising spin
  glasses.
\end{itemize}

A convenient way to capture the binary choice of each agent is to encode it by an Ising spin $s_i \in \{+1,-1\}$, where $s_i = +1$ corresponds to $\ell_i^{(1)}$ and $s_i = -1$ to $\ell_i^{(2)}$.
With this encoding the occupancy \(n_l\) becomes a linear function of the spins
\(s_i\), and the cost function \(E\) can be rewritten exactly as a quadratic form. The
resulting expression has the same structure as an Ising Hamiltonian with quenched
random couplings and random fields. Therefore, the computational problem of finding the
agent distribution that minimises \(E\) is equivalent to locating the ground state of a
spin glass: a problem that is generically NP‑hard and whose energy landscape contains a
multitude of local minima \cite{mezard1987spin, Kirkpatrick1983}.

Earlier numerical studies of the PMG observed that even the most efficient stochastic
strategies often become trapped in frozen, sub‑optimal configurations \cite{vemula2025}. These frozen
states are precisely local minima of the effective energy landscape: agents cannot
unilaterally flip their choices to improve their minority status, yet the overall
cost function remains far from its global minimum : a hallmark of spin‑glass behaviour. The
mapping we develop in the next section makes this connection rigorous.

\section{Encoding Agent Choices as Spins}

For each agent $i$ we introduce an Ising spin variable $s_i \in \{+1,-1\}$:
\[
s_i = \begin{cases}
+1 & \text{if agent } i \text{ chooses } \ell_i^{(1)},\\
-1 & \text{if agent } i \text{ chooses } \ell_i^{(2)} .
\end{cases}
\]
The occupancy of location $l$ can then be written as
\begin{equation}
n_l = \sum_{i=1}^{N} \left[ \delta_{\ell_i^{(1)},l}\; \frac{1+s_i}{2} + \delta_{\ell_i^{(2)},l}\; \frac{1-s_i}{2} \right].
\label{eq:nl}
\end{equation}

To separate the spin‑dependent part, define binary coefficients
\[
a_i^{(l)} = \delta_{\ell_i^{(1)},l}, \qquad b_i^{(l)} = \delta_{\ell_i^{(2)},l}.
\]
Then Eq.~(\ref{eq:nl}) becomes
\begin{equation}
n_l = \sum_{i=1}^{N} \left[ \frac{a_i^{(l)} + b_i^{(l)}}{2} + \frac{a_i^{(l)} - b_i^{(l)}}{2}\, s_i \right] 
= C_l + \sum_{i=1}^{N} d_i^{(l)} s_i ,
\label{eq:nl_compact}
\end{equation}
where
\begin{equation}
C_l = \sum_{i=1}^{N} \frac{a_i^{(l)} + b_i^{(l)}}{2}, \quad
d_i^{(l)} = \frac{a_i^{(l)} - b_i^{(l)}}{2} \in \{-1/2, 0, +1/2\}.
\end{equation}
The constant $C_l$ counts the average number of agents that list location $l$ among their two choices, while $d_i^{(l)}$ is non‑zero only when agent $i$ has distinct choices and one of them is $l$.


By inserting the compact form (3) into the cost function (1), we get:
\begin{equation}
E = \sum_{l=1}^{D} \left( C_l - \frac{N}{D} + \sum_i d_i^{(l)} s_i \right)^2 .
\label{eq:5}
\end{equation}

Expanding the square gives
\begin{widetext}
\begin{equation}
\begin{aligned}
E &= \sum_{l=1}^{D} \Bigg[ \left( \sum_i d_i^{(l)} s_i \right)^2 
      + 2 \left( C_l - \frac{N}{D} \right) \sum_i d_i^{(l)} s_i 
      + \left( C_l - \frac{N}{D} \right)^2 \Bigg] \\
  &= \sum_{i,j} s_i s_j \sum_l d_i^{(l)} d_j^{(l)}
     + 2 \sum_i s_i \sum_l \left( C_l - \frac{N}{D} \right) d_i^{(l)}
     + \text{const.}
\label{eq:6}
\end{aligned}
\end{equation}
\end{widetext}

Because $s_i^2 = 1$, the diagonal terms $i=j$ in the first sum contribute only a constant shift. The cost function then takes the form of an Ising spin glass Hamiltonian:
\begin{equation}
\ H = \sum_{i,j} J_{ij} s_i s_j + \sum_i h_i s_i + \text{const} ,
\label{eq:spin_glass}
\end{equation}
where the couplings and fields are given by
\begin{equation}
J_{ij} = \sum_{l=1}^{D} d_i^{(l)} d_j^{(l)}, \qquad
h_i = 2 \sum_{l=1}^{D} \left( C_l - \frac{N}{D} \right) d_i^{(l)} .
\label{eq:Jh}
\end{equation}

\begin{figure}[ht]
\centering
\includegraphics[width=0.9\linewidth]{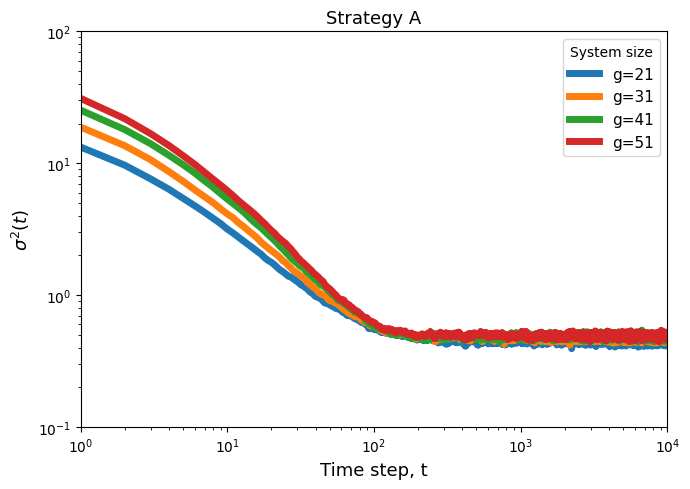}
\includegraphics[width=0.9\linewidth]{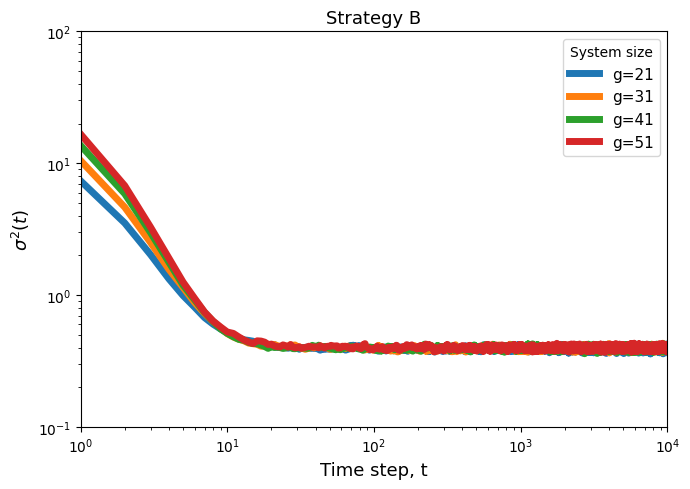}
\includegraphics[width=0.9\linewidth]{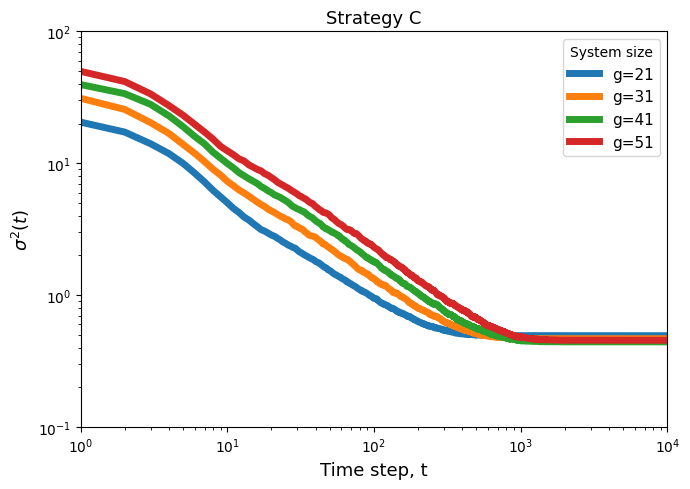}
\caption{Variance evolution for D=101 and different values of $g$ across time, for various strategies: A , B and C. The variance decreases initially and then stabilizes.}
\label{fig:variance}
\end{figure}

\begin{figure}[ht]\centering
\includegraphics[width=0.9\linewidth]{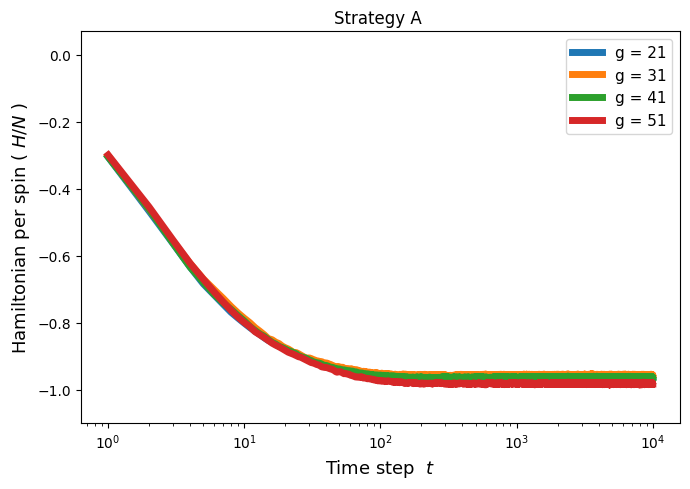}
\includegraphics[width=0.9\linewidth]{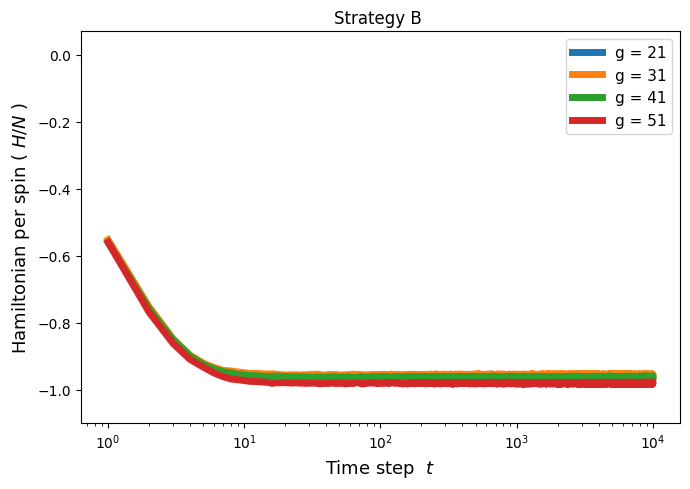}
\includegraphics[width=0.9\linewidth]{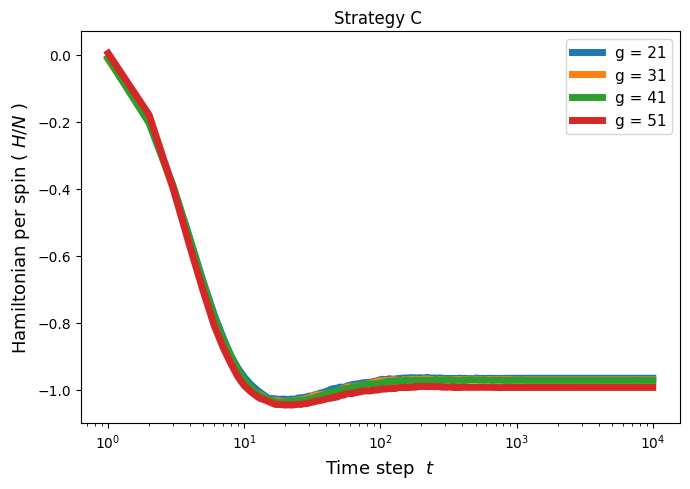}
\caption{Time evolution of the Hamiltonian (Eg. \ref{eq:spin_glass}) for D=101 and  different values of $g$, various strategies: A , B and C. The Hamiltonian decreases and reaches a saturation indicating convergence to local minima of the spin energy landscape.}
\label{fig:hamiltonian}
\end{figure}

\vspace{12mm}

We will henceforth refer to $H$ as the energy of the spin glass. The values of $J_{ij}$ can be $0, \pm 1/4, \pm 1/2$, depending on how the agents' allowed location pairs overlap. Mixed signs generate frustration, exactly as in spin glass models. The random fields $h_i$ reflect the global imbalance created by the fixed background occupancy $C_l$. Minimizing $H$ is therefore equivalent to finding the ground state of this Ising spin glass, which in turn yields the agent distribution with the lowest population variance.

\section{Results}
The optimization of the population distribution in the PMG have been attempted in several different stochastic strategies \cite{vemula2025, biswas2026}. The strategies are generally crowd avoiding in nature and uses the information of the population in the current choice of an agent and the population at their alternate choices. There are three strategies that we discuss here. In all of these cases, an agent ($i$-th agent, say) switches to their alternate location ($y$) with a probability when the population at their current location ($x$) at a time $t$ (say, $n_i^x(t)$) exceeds their expected population $\mathcal{E_i}(t)$. The definition of this expected population differs in the different strategies. However, the switching probability is such that the current population changes towards the expected population:
\begin{equation}
    p_i(t)=\frac{n_i^x(t)-\mathcal{E_i}(t)}{2n_i^x(t)},
\end{equation}
provided, of course $\mathcal{E_i}(t)<n_i^x(t)$, otherwise $p_i(t)=0$. Now, the form of the expected population is as follows: 

\begin{itemize}
    \item For strategy A: $\mathcal{E_i}(t)=g$, where $g$ is the average population,
    \item For strategy B: $\mathcal{E_i}(t)=n_i^y(t)$, where $n_i^y(t)$ is the population at the alternate location ($y$) for the $i$-th agent at that instant,
    \item For strategy C: $\mathcal{E_i}(t)=n_i^y(t^{\prime})$, where $t^{\prime}$ is the time where the $i$-th agent had last vitited their alternate location (of course $t^{\prime}<t$). 
\end{itemize}

It was shown before \cite{vemula2025} that the Strategy C, having a delayed information of the alternate population, performed the best in terms of reducing the population variance. 

Here, we look at the same strategies, but along with the population variance, we also look at the energy of the spin glass Hamiltonian, after the system has reached a stable state. The population variances are shown in Fig. \ref{fig:variance} for different strategies and different $g$ values in each strategy. As mentioned before, the strategy C, having a delayed information of crowd, is able to avoid overcrowding the best (hence reaches a lower variance), even though its dynamics take a longer time to saturate as compared to the other strategies. In Fig. \ref{fig:hamiltonian}, the corresponding energy per spin are plotted, while the population dynamics was following the different strategies mentioned above. The  saturation values of the energies are lower than what is expected for the standard SK spin glass (see e.g., \cite{mezard1987spin}), which is not entirely unexpected, given the random fields are present here. Although, it is important to note that the dynamics of the model here do not follow any of the energy minimization techniques followed in SK spin glasses, but simply obtained from the mapping with the population dynamics. 

The other surprising point is the saturation of the energy values for strategy C at a much smaller time scale (having only a weak dependence with system size), than the population variance. It shows that the overall energy function does not put all population fluctuations in equal footing i.e., larger population fluctuation can still lead to smaller energy values. The physical mechanism behind this behavior still needs to be explored. 
\vspace{6mm}
\section{Discussion and Conclusion}

We have mapped the problem of minimizing population variance in the parallel minority game to finding the ground state of an Ising spin glass. The quenched couplings $J_{ij}$ and fields $h_i$ encode the fixed alternatives of each agent, and the frustration inherent in overlapping choices manifests as mixed‑sign interactions. This mapping explains the glassy, frozen states observed in stochastic strategies and provides a rigorous connection between the PMG and the well developed theory of disordered systems.

The spin‑glass perspective not only clarifies the origin of the computational hardness of the PMG but also opens the door to using powerful optimization tools from spin‑glass theory such as simulated annealing, replica methods, and mean‑field algorithms (see e.g., \cite{sg_pre}) to systematically search for agent distributions that minimize the global variance. This connection establishes a new foundation for both theoretical analysis and practical optimization of parallel resource allocation problems.

However, it also needs to be noted that ground state search algorithms of the spin glass systems necessarily requires the full structure of the Hamiltonian in order to have a spin update. But in terms of the PMG, that implies sharing the information of population of all locations to all agents. This is a significant increase in the population information, hence it is expected to manifest a much reduced population variance for the PMG systems.

\bibliography{references}

@article{chakraborti2011econophysicsII,
  title        = {Econophysics review: II. Agent‐based models},
  author       = {Chakraborti, Anirban and Muni Toke, Ioane and Patriarca, Marco and Abergel, Fr{\'e}d{\'e}ric},
  journal      = {Quantitative Finance},
  volume       = {11},
  number       = {7},
  pages        = {1013--1041},
  year         = {2011},
  doi          = {10.1080/14697688.2010.539249},
  publisher    = {Taylor \& Francis}
}

@article{challet1997,
  author  = {Challet, D. and Zhang, Y.-C.},
  title   = {Emergence of cooperation and organization in an evolutionary game},
  journal = {Physica A},
  volume  = {246},
  pages   = {407--418},
  year    = {1997},
  doi     = {10.1016/S0378-4371(97)00295-6}
}

@article{challet1998,
  author  = {Challet, D. and Zhang, Y.-C.},
  title   = {On the minority game: Analytical and numerical studies},
  journal = {Physica A},
  volume  = {256},
  pages   = {514--532},
  year    = {1998},
  doi     = {10.1016/S0378-4371(98)00470-7}
}

@article{challet1999,
  author  = {Challet, D. and Marsili, M.},
  title   = {Phase transition and symmetry breaking in the minority game},
  journal = {Phys. Rev. E},
  volume  = {60},
  pages   = {R6271--R6274},
  year    = {1999},
  doi     = {10.1103/PhysRevE.60.R6271}
}

@article{chakraborti2015,
  author  = {Chakraborti, A. and Challet, D. and Chatterjee, A. and Marsili, M. and Zhang, Y.-C. and Chakrabarti, B. K.},
  title   = {Statistical mechanics of competitive resource allocation using agent-based models},
  journal = {Phys. Rep.},
  volume  = {552},
  pages   = {1--25},
  year    = {2015},
  doi     = {10.1016/j.physrep.2014.09.004}
}

@article{challet2001local,
  author  = {Challet, D. and Chessa, A. and Marsili, M. and Zhang, Y.-C.},
  title   = {From minority games to real markets},
  journal = {Quant. Finance},
  volume  = {1},
  pages   = {168--176},
  year    = {2001},
  doi     = {10.1088/1469-7688/1/2/303}
}

@article{dhar2011,
  author  = {Dhar, D. and Sasidevan, V. and Chakrabarti, B. K.},
  title   = {Emergent cooperation amongst competing agents in minority games},
  journal = {Physica A},
  volume  = {390},
  pages   = {3477--3485},
  year    = {2011},
  doi     = {10.1016/j.physa.2011.05.040}
}

@article{chakrabarti2009,
  author  = {Chakrabarti, A. S. and Chakrabarti, B. K. and Chatterjee, A. and Mitra, M.},
  title   = {The Kolkata Paise Restaurant problem and resource utilization},
  journal = {Physica A},
  volume  = {388},
  pages   = {2420--2426},
  year    = {2009},
  doi     = {10.1016/j.physa.2009.02.037}
}

@article{ghosh2010,
  author  = {Ghosh, A. and Chatterjee, A. and Chakrabarti, B. K. and Mitra, M.},
  title   = {Statistics of the Kolkata Paise Restaurant problem},
  journal = {New J. Phys.},
  volume  = {12},
  pages   = {075073},
  year    = {2010},
  doi     = {10.1088/1367-2630/12/7/075073}
}

@article{biswas2012,
  author  = {Biswas, S. and Ghosh, A. and Chatterjee, A. and Naskar, T. and Chakrabarti, B. K.},
  title   = {Continuous transition of social efficiencies in the stochastic-strategy minority game},
  journal = {Phys. Rev. E},
  volume  = {85},
  pages   = {031104},
  year    = {2012},
  doi     = {10.1103/PhysRevE.85.031104}
}

@article{biswas2021,
  author  = {Biswas, S. and Mandal, A. K.},
  title   = {Parallel minority game and its application in movement optimization during an epidemic},
  journal = {Physica A},
  volume  = {561},
  pages   = {125271},
  year    = {2021},
  doi     = {10.1016/j.physa.2020.125271}
}

@article{sysiaho2004adaptive,
  author  = {Sysi-Aho, M. and Chakraborti, A. and Kaski, K.},
  title   = {Searching for good strategies in adaptive minority games},
  journal = {Phys. Rev. E},
  volume  = {69},
  pages   = {036125},
  year    = {2004},
  doi     = {10.1103/PhysRevE.69.036125}
}

@article{vemula2025,
title = {Efficient strategy for Parallel Minority Games},
journal = {Physica A: Statistical Mechanics and its Applications},
volume = {688},
pages = {131373},
year = {2026},
issn = {0378-4371},
doi = {https://doi.org/10.1016/j.physa.2026.131373},
url = {https://www.sciencedirect.com/science/article/pii/S0378437126001093},
author = {Ankith Reddy Vemula and Soumyajyoti Biswas},
}

@article{pmg_vac,
	author={Xue, Chenli and Luo, Xiaofeng and Sun, Gui-Quan},
	title={Vaccination-Transmission Coupled Mechanism Based on Parallel Minority Game},
	journal={Chinese Physics B},
	url={http://iopscience.iop.org/article/10.1088/1674-1056/ae306a},
	year={2025}
}

@article{sg_pre,
  title = {Classical annealing of the {S}herrington-{K}irkpatrick spin glass using {S}uzuki-{K}ubo mean-field Ising dynamics},
  author = {Das, Soumyaditya and Biswas, Soumyajyoti and Chakrabarti, Bikas K.},
  journal = {Phys. Rev. E},
  volume = {112},
  issue = {1},
  pages = {014104},
  numpages = {7},
  year = {2025},
  month = {Jul},
  publisher = {American Physical Society},
  doi = {10.1103/www8-3ts1},
  url = {https://link.aps.org/doi/10.1103/www8-3ts1}
}

@article{sk_model,
  title = {Solvable Model of a Spin-Glass},
  author = {Sherrington, David and Kirkpatrick, Scott},
  journal = {Phys. Rev. Lett.},
  volume = {35},
  issue = {26},
  pages = {1792--1796},
  numpages = {0},
  year = {1975},
  month = {Dec},
  publisher = {American Physical Society},
  doi = {10.1103/PhysRevLett.35.1792},
  url = {https://link.aps.org/doi/10.1103/PhysRevLett.35.1792}
}

@misc{tyagi2026,
      title={Active-Absorbing Phase Transitions in the Parallel Minority Game}, 
      author={Aryan Tyagi and Soumyajyoti Biswas and Anirban Chakraborti},
      year={2026},
      eprint={2512.22826},
      archivePrefix={arXiv},
      primaryClass={cond-mat.stat-mech},
      url={https://arxiv.org/abs/2512.22826}, 
}

@book{mezard1987spin,
  title={Spin Glass Theory And Beyond: An Introduction To The Replica Method And Its Applications},
  author={Mezard, M. and Parisi, G. and Virasoro, M.A. and Hopfield, J.J.},
  isbn={9789813103917},
  series={World Scientific Lecture Notes In Physics},
  url={https://books.google.co.in/books?id=DwY8DQAAQBAJ},
  year={1987},
  publisher={World Scientific Publishing Company}
}

@article{
Kirkpatrick1983,
author = {S. Kirkpatrick  and C. D. Gelatt  and M. P. Vecchi },
title = {Optimization by Simulated Annealing},
journal = {Science},
volume = {220},
number = {4598},
pages = {671-680},
year = {1983},
doi = {10.1126/science.220.4598.671},
URL = {https://www.science.org/doi/abs/10.1126/science.220.4598.671},
}

@misc{biswas2026,
      title={Local and global optimization in Parallel Minority Games}, 
      author={Soumyajyoti Biswas and Jnanesh Yaramati and Kavya Bellamkonda and Krishna Rastogi and Devesh Chaudhary},
      year={2026},
      eprint={2605.05141},
      archivePrefix={arXiv},
      primaryClass={physics.soc-ph},
      url={https://arxiv.org/abs/2605.05141}, 
}

\end{document}